\documentclass[twocolumn,twocolappendix]{aastex631}
\usepackage{soul}

\newcommand{\jh}[1]{{\color{black} #1}}

\shorttitle{JVLA Observations of CIDA 1}
\shortauthors{Hashimoto et al.}

\graphicspath{{./}{figures/}}

\begin{document}

\title{Centimeter-sized Grains in the Compact Dust Ring around Very Low Mass Star CIDA~1}

\correspondingauthor{Jun Hashimoto}
\email{jun.hashimto@nao.ac.jp}

\author[0000-0002-3053-3575]{Jun Hashimoto}
\affil{Astrobiology Center, National Institutes of Natural Sciences, 2-21-1 Osawa, Mitaka, Tokyo 181-8588, Japan}
\affil{Subaru Telescope, National Astronomical Observatory of Japan, Mitaka, Tokyo 181-8588, Japan}
\affil{Department of Astronomy, School of Science, Graduate University for Advanced Studies (SOKENDAI), Mitaka, Tokyo 181-8588, Japan}

\author[0000-0003-2300-2626]{Hauyu Baobab Liu}
\affiliation{Department of Physics, National Sun Yat-Sen University, No. 70, Lien-Hai Road, Kaohsiung City 80424, Taiwan, R.O.C.}

\author[0000-0001-9290-7846]{Ruobing Dong}
\affil{Department of Physics \& Astronomy, University of Victoria, Victoria, BC, V8P 5C2, Canada}

\author[0000-0001-5830-3619]{Beibei Liu}
\affil{Institute for Astronomy, School of Physics, Zhejiang University, 38 Zheda Road, Hangzhou 310027, China}

\author{Takayuki Muto}
\affil{Division of Liberal Arts, Kogakuin University, 1-24-2, Nishi-Shinjuku, Shinjuku-ku, Tokyo 163-8677, Japan}
\affil{Leiden Observatory, Leiden University, P.O. Box 9513, NL-2300 RA Leiden, The Netherlands}
\affil{Department of Earth and Planetary Sciences, Tokyo Institute of Technology, 2-12-1 Oh-okayama, Meguro-ku, Tokyo 152-8551, Japan}

\author[0000-0003-2887-6381]{Yuka Terada}
\affil{Institute of Astronomy and Astrophysics, Academia Sinica, 11F of Astronomy-Mathematics Building, AS/NTU No.1, Sec. 4,
Roosevelt Rd, Taipei 10617, Taiwan, ROC}
\affil{Department of Astrophysics, National Taiwan University, Taipei 10617, Taiwan, R.O.C.} 

\begin{abstract}
We examined the grain size in the dust ring encircling the 0.19~$M_\sun$ T Tauri star CIDA 1 using the Karl G. Jansky Very Large Array (JVLA) at multiple centimeter wavelengths, with a spatial resolution of 0$\farcs$2--0$\farcs$9. We detected distinct partial-ring structures at these wavelengths around CIDA~1. Based on spatial distributions and spectral indexes, we determined that these centimeter emissions originated from dust, rather than free-free or synchrotron emissions. To estimate the maximum grain size ($a_{\rm max}$) within the ring, we compared the observed spectral energy distribution (SED) with SEDs calculated for different $a_{\rm max}$ values using radiative transfer calculations. Our findings indicate an $a_{\rm max}$ value of approximately 2.5~cm in the ring, assuming the dust opacity can be approximated by the DSHARP models. These results suggest that grain growth took place within the CIDA~1 ring, potentially facilitating more efficient planet formation through pebble accretion scenarios involving centimeter-sized pebbles.
\end{abstract}


\section{Introduction} \label{sec:intro}
The initial stage of planet formation in protoplanetary disks is grain growth. Understanding this process and determining the maximum grain size are essential because subsequent planet formation mechanisms such as streaming instability \citep[e.g.,][]{joha2014a} and pebble accretion \citep[e.g.,][]{Liu2020pebbleaccretion} depend on them \citep{Carrera2015,Morbidelli2015,Ormel2018,Drazkowska2022}.

Two methods exist for determining the sizes of dust particles based on millimeter/centimeter (mm/cm) continuum emission. The first approach involves observing polarized dust continuum emission in the mm/cm range, resulting from scattering \citep{Kataoka2015a}. The polarization fraction in the dust emission reaches its maximum when the dust grains grow to a size of approximately $a_{\rm max} \sim \lambda/2\pi$, where $\lambda$ represents the observing wavelength. This method requires a signal-to-noise ratio (SNR) of approximately 100--1,000 in Stokes~$I$ and is well-suited for studying bright objects like DSHARP disks \citep{andr2018}. The second approach involves constructing the spectral energy distribution (SED) \citep[e.g.,][] {Testi2003CQTau,Perez2012AS209,Testi2014ppvi,Perez2015CYTau,Tazzari2016GrainGrowth,Liu2019SED,Ueda2020TWHya,Ueda2021HLTau}. This method relies on the fact that grains emit thermal radiation most efficiently at wavelengths similar to their sizes \citep[e.g.,][]{Draine2006}. By analyzing the SED, information about the grain sizes can be inferred.

Polarization observations in the (sub-)millimeter range have provided insights into the maximum grain size ($a_{\rm max}$) in protoplanetary disks, estimated to be around 100~$\mu$m \citep[e.g.,][]{Kataoka2016a,Kataoka2016b,Ohashi2018a,Ohashi2020,Bacciotti2018,Dent2019}. The efficiency of pebble accretion, crucial for planet growth, exhibits a size dependency \citep{Liu2018,Ormel2018}. When accreting pebbles of 100~$\mu$m, planet growth generally occurs at a slower pace compared to mm to cm-sized pebbles \citep[e.g.,][]{Morbidelli2015}. Consequently, the formation of giant planets with sub-mm dust grains may require longer timescales, potentially exceeding the typical disk lifetime of a few million years \citep{Mamajek2009}. Thus the grain-size problem remains a prominent topic in the field of planet formation. However, it is important to note that ALMA polarization observations have exhibited a bias towards bright disks, leaving uncertainties regarding the ubiquity of dust grains with sub-mm $a_{\rm max}$ in disks.  

Some objects have indeed exhibited the presence of millimeter-/centimeter-sized grains within their disks \citep[e.g.,][]{Carrasco-Gonzalez2019HLTau,Ueda2021HLTau,Macias2021TWHya,Hashimoto2022,Zhang2023HLTau}. In this study, we present another example: CIDA~1 (2MASS~J04141760$+$2806096), which contains dust grains of millimeter to centimeter sizes within its disk. CIDA~1, located at $d=134.6$~pc in the Taurus star-forming region, has a mass of 0.19~$M_\sun$ and an effective temperature of 3,197~K \citep{Kurtovic2021a,gaia2016,gaia2022}. It has been extensively observed with ALMA in bands~3, 4, 6, and 7 \citep{Ricci2014a,Simon2017,Pinilla2018CIDA1,Pinilla2021a,Kurtovic2021a}, revealing a ring/gap structure at a radius of around 20~au in the dust continuum image at bands~4 and 7 \citep{Pinilla2021a}. Recent hydrodynamic simulations indicate that a giant planet with a minimum mass of 1.4~$M_{\rm Jup}$ could account for the observed ring/gap structure around CIDA~1 \citep{Curone2022}. The spectral index $\alpha_{\rm band4\mathchar`-7}$ between ALMA bands~4 and 7 is measured to be 2.0~$\pm$~0.2, and the continuum ring at bands~4 and 7 is optically thin \citep{Pinilla2021a}, suggesting that the dust grains within the CIDA~1 ring have grown to sizes on the order of millimeters to centimeters \citep[e.g.,][]{Draine2006}. 

\section{Observations and results} \label{sec:obs}

The JVLA observations \jh{with B configuration} were conducted as part of program ID 23A-124 (PI: J. Hashimoto), and the details are provided in Table 1. The data were calibrated using the Common Astronomy Software Applications (CASA) package \citep{mcmu07}, following the calibration scripts provided by JVLA. Due to weak emission from the object, self-calibration of the visibilities was not performed. The stellar position was corrected by proper motion (8.285, -23.607) mas/yr \citep{gaia2016,gaia2022}. The corrected ICRS coordinate for CIDA~1 is (4h 14m 17.62395s, 28d 6m 9.11174s). Subsequently, \jh{multi-term multi-frequency synthesis imaging with {\tt nterm}~=~2 \citep{Rau2011MultiscaleClean}} was performed using the CASA {\tt tclean} task, as summarized in Table~\ref{tab:image}. \jh{We have attempted phase-only self-calibration on the Ka-band data, which has the highest signal-to-noise ratio (SNR); see below. However approximately 62\% of solutions on the solution intervals of 100s, 300s, 1,000s were flagged with a minimum SNR of 2.5. Thus we have decided not to apply self-calibration.}

\begin{deluxetable}{lc}[htbp]
\tablewidth{0pt} 
\tablecaption{JVLA observations \jh{with B configuration} \label{tab:obs}}
\tablehead{
\colhead{Observations}      & \colhead{}    
}
\startdata
Observing date (UT)         & 2023.Jan.13, 14, 15, 20,   \\
                            & Feb.3, Apr.16 \\
Project code                & 23A-124 (PI: J. Hashimoto)       \\
Central frequency (GHz)     & 44 (Q), 33 (Ka), \\
                            & 15 (Ku), 10 (X) \\
Continuum band width (GHz)  & 8 (Q, Ka), 6 (Ku), 4 (X) \\
Time on source (min)        & 48.5 (Q), 156.0 (Ka),  \\
                            & 20.7 (Ku), 6.2 (X) \\
Number of antennas          & 27 (Jan.13, 14, 20, Feb.3),  \\
                            & 26 (Jan.15, Apr.16) \\
Baseline lengths (km)       & 0.243 to 11.1                  \\
Bandpass calibrator         & J0319$+$4130                     \\
Flux calibrator             & 3C147                     \\
Phase calibrator            & J0403$+$2600                   \\
\enddata
\end{deluxetable}

\begin{deluxetable*}{lcccccccc}[htbp]
\tabletypesize{\footnotesize}
\tablewidth{0pt} 
\tablecaption{Imaging parameters\label{tab:image}}
\tablehead{
\colhead{}      & \colhead{X band (10 GHz)} & \multicolumn{3}{c}{Ku band (15 GHz)} & \multicolumn{3}{c}{Ka band (33 GHz)} & \colhead{Q band (44 GHz)}    
}
\startdata
\multicolumn{9}{c}{For imaging in Figure~\ref{fig:JVLAimg} without splitting observing bands} \\
Robust clean parameter & 2.0  & \multicolumn{3}{c}{2.0}  & \multicolumn{3}{c}{0.0} & 2.0 \\
Beam shape             & \jh{0$\farcs$89$\times$0$\farcs$74} & \multicolumn{3}{c}{\jh{0$\farcs$58$\times$0$\farcs$51}} & \multicolumn{3}{c}{\jh{0$\farcs$17$\times$0$\farcs$16}} & \jh{0$\farcs$20$\times$0$\farcs$18}\\
r.m.s. noise ($\mu$Jy/beam) & 11.7 & \multicolumn{3}{c}{\jh{4.5}} & \multicolumn{3}{c}{5.4} & \jh{18.3} \\ \hline
\multicolumn{9}{c}{For measuring flux in Table~\ref{tab:fluxdensity} with \jh{or without} splitting observing bands} \\
Split frequency (GHz)  & \jh{8--12} & 12--15 & 15--18 & \jh{12--18} & 29--33 & 33--37 & \jh{29--37} & \jh{40--48} \\
                       &\jh{(no split)}&\jh{(split)}&\jh{(split)}&\jh{(no split)}&\jh{(split)}&\jh{(split)}&\jh{(no split)}&\jh{(no split)}\\
Robust clean parameter & 2.0 & 2.0 & 2.0 & 2.0 & 2.0 & 2.0 & 2.0 & 2.0 \\
Beam shape             & \jh{0$\farcs$89$\times$0$\farcs$74} & \jh{0$\farcs$63$\times$0$\farcs$55} & \jh{0$\farcs$54$\times$0$\farcs$46} & \jh{0$\farcs$58$\times$0$\farcs$51} & \jh{0$\farcs$29$\times$0$\farcs$25} & \jh{0$\farcs$26$\times$0$\farcs$22} & \jh{0$\farcs$27$\times$0$\farcs$23} & \jh{0$\farcs$20$\times$0$\farcs$18}\\
r.m.s. noise ($\mu$Jy/beam) & 11.7 & 6.1 & \jh{6.7} & \jh{4.5} & \jh{5.4} & 5.2 & \jh{3.7} & \jh{18.3} \\ \hline
\enddata
\end{deluxetable*}

Figure 1 showcases the JVLA images obtained at Q~band (40--48~GHz), Ka~band (29--37~GHz), Ku~band (12--18~GHz), and X~band (8--12~GHz). To achieve a comparable spatial resolution of 0\farcs18 to the Q-band image, we utilized Briggs Robust~$=$~0.0 for the Ka-band~image. Our observations revealed significant signals at Q, Ka, and Ku bands, with peak flux densities of \jh{87.6 (4.8~$\sigma$), 36.9 (6.8~$\sigma$), and 19.0 (4.2~$\sigma$) $\mu$Jy/beam}, respectively. The integrated flux measurements, exceeding 3~$\sigma$, are presented in Table~\ref{tab:fluxdensity}. To improve the signal-to-noise ratio (SNR), Briggs Robust~$=$~2.0 was applied for the Ka-band flux (refer to Table~\ref{tab:image}). As the beam size of approximately 0\farcs4 is similar to the size of the CIDA~1 ring, the peak flux density from the Ku-band image was utilized as the total flux.

\begin{figure*}[htbp]
         \includegraphics[width=\linewidth]{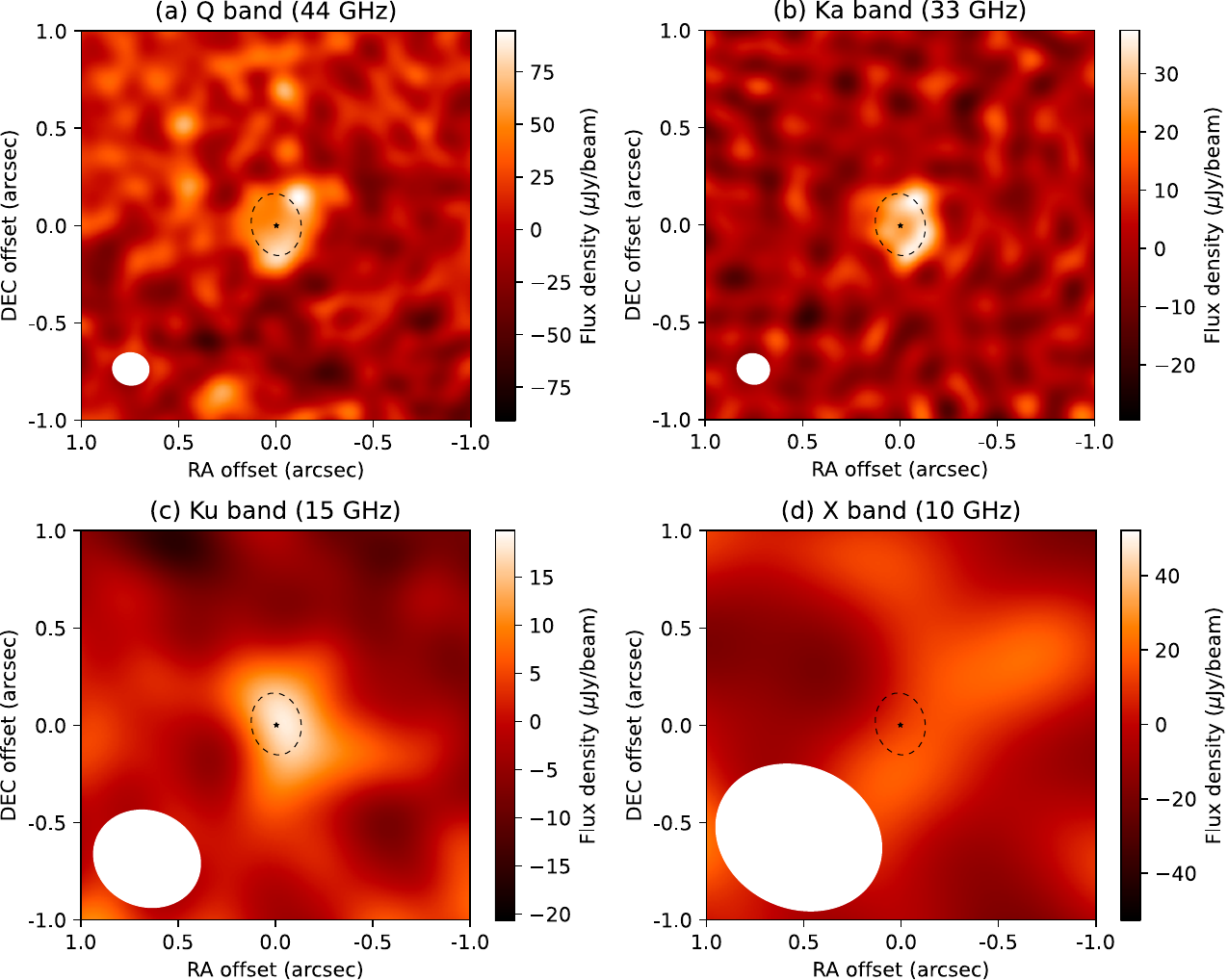} 
    \caption{
    The JVLA images (color) show the CIDA 1 ring overlaid with the ALMA band-7 data (dashed ellipses; $r=0\farcs16$; $i=37.4$~deg; P.A.~$=10.8$~deg; \citealt{Pinilla2021a}). Panels (a) to (d) depict the Q-band (40--48~GHz), Ka-band (29--37~GHz), Ku-band (12--18~GHz), and X-band (8-12~GHz) images, respectively. The stellar position is marked by a black star, corresponding to the ICRS coordinate of (4h 14m 17.62395s, 28d 6m 9.11174s). The proper motion (pmRA~$=8.285$~mas/yr, pmDEC~$=-23.607$~mas/yr; \citealp{gaia2016,gaia2022}) has been corrected. The synthesized beams are illustrated as white ellipses in the lower left corner.
    }
    \label{fig:JVLAimg}
\end{figure*}

The signals at Q- and Ka-bands are clearly spatially resolved. The locations of strong signals, depicted in white in Figure~\ref{fig:JVLAimg}, well trace the ring structure detected by ALMA sub-millimeter emission \citep{Pinilla2021a} as indicated by the dashed black ellipse. Although the ring appears asymmetric at Q- and Ka-bands, with a brighter region in the northwest, the contrast between the peak position and the opposing side within the ring are measured to $1.40\pm0.43$ and $1.61\pm0.40$ at Q- and Ka-bands, respectively. The observed asymmetry at the Ka-band might be genuine. However, due to the low SNR in the contrast, we refrain from further discussing the asymmetry. It is imperative to conduct observations with higher S/N ratios to investigate potential asymmetries within the CIDA~1 ring.

\begin{figure*}[htbp]
         \includegraphics[width=\linewidth]{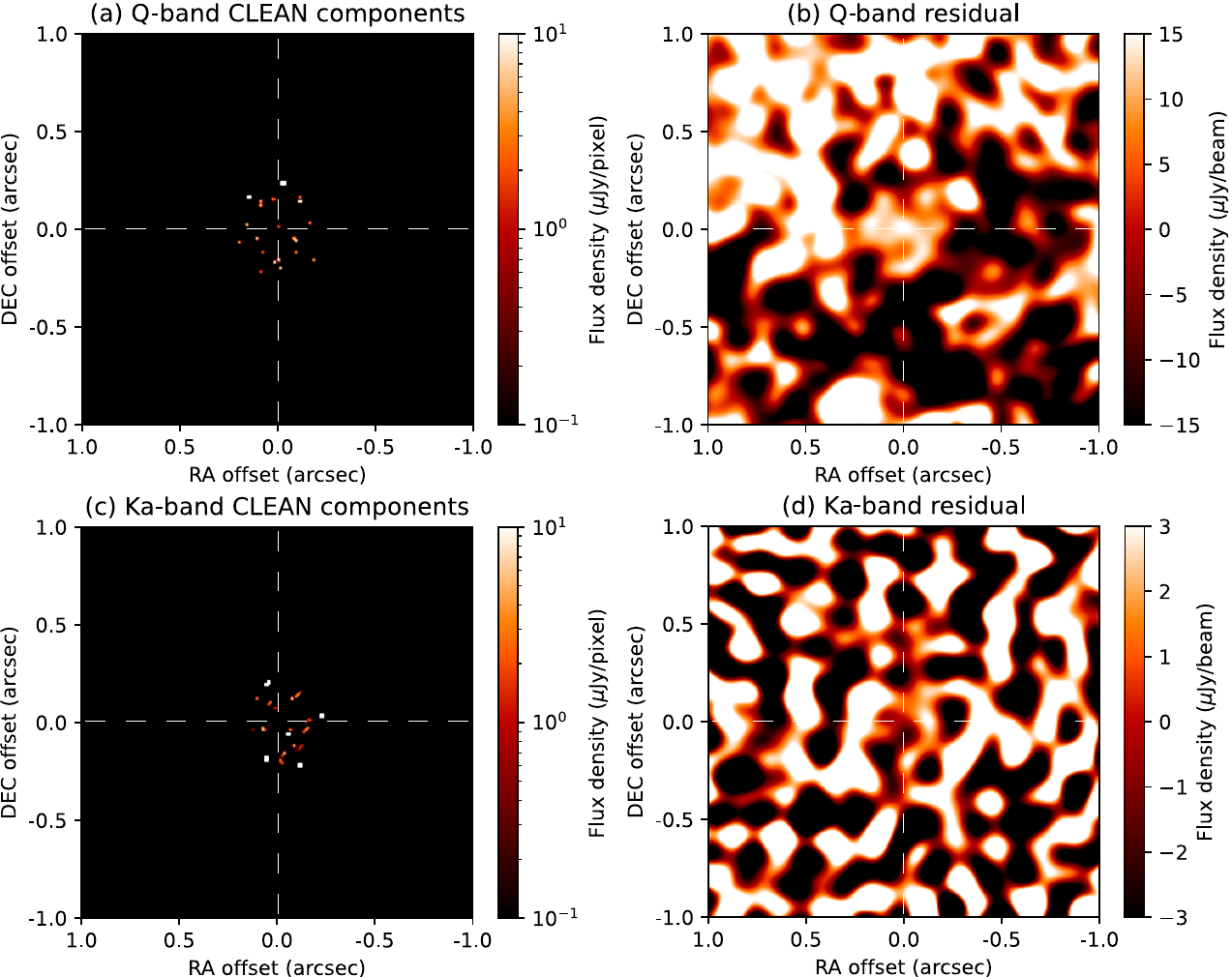} 
    \caption{
    \jh{The inner disk component is explored in Q- and Ka-band data. Left: the CLEAN components. There is a CLEAN component with 4.2~$\mu$Jy/pixel in the central region of Q-band data in panel~(a). Right: residual images with a specific mask on the ring region indicated by the dashed ellipse in Figure~\ref{fig:JVLAimg}. The peak flux of the central region in both residual images is less than 1~$\sigma$ (Table~\ref{tab:image}). }
    }
    \label{fig:innerdisk}
\end{figure*}

\jh{The previously detected central point source in ALMA band-4 and 7 images \citep{Pinilla2021a} is not robustly recovered in our JVLA observations (Figure~\ref{fig:JVLAimg}), possibly due to the lower spatial resolution. To investigate the central source in Q- and Ka-band data, we first checked their CLEAN components. Figures~\ref{fig:innerdisk}(a) and (c) show the CLEAN components of Q- and Ka-band data. There is a CLEAN component with 4.2~$\mu$Jy/pixel in the central region of the Q-band model. We then performed {\tt tclean} again, with a mask on the ring region indicated by the dashed ellipse in Figure~\ref{fig:JVLAimg}. Figures~\ref{fig:innerdisk}(b) and (d) show the resulting Q- and Ka-band residual images. We find that the peak flux of the central source in both residual images is less than 1~$\sigma$ (Table~\ref{tab:image}). High resolution and high sensitivity observations are needed for further investigations of the central point source.} 

In order to perform SED modeling of the ring in \S~\ref{sec:amax}, it is necessary to estimate the flux contribution of the central source to the integrated flux. The visibility fitting analysis of \citet{Pinilla2021a} using ALMA band-4 and 7 data indicated that the size (FWHM) of the central source is comparable to or smaller than the beam size of approximately 30-50~mas, suggesting that the central source is not well spatially resolved in the ALMA images. Therefore, we used the peak flux density of the central source, which is summarized in Table~\ref{tab:fluxdensity}.

\begin{figure*}[htbp]
\centering
    \begin{tabular}{ ll }
         \hspace{-0.cm}\includegraphics[height=8.5cm]{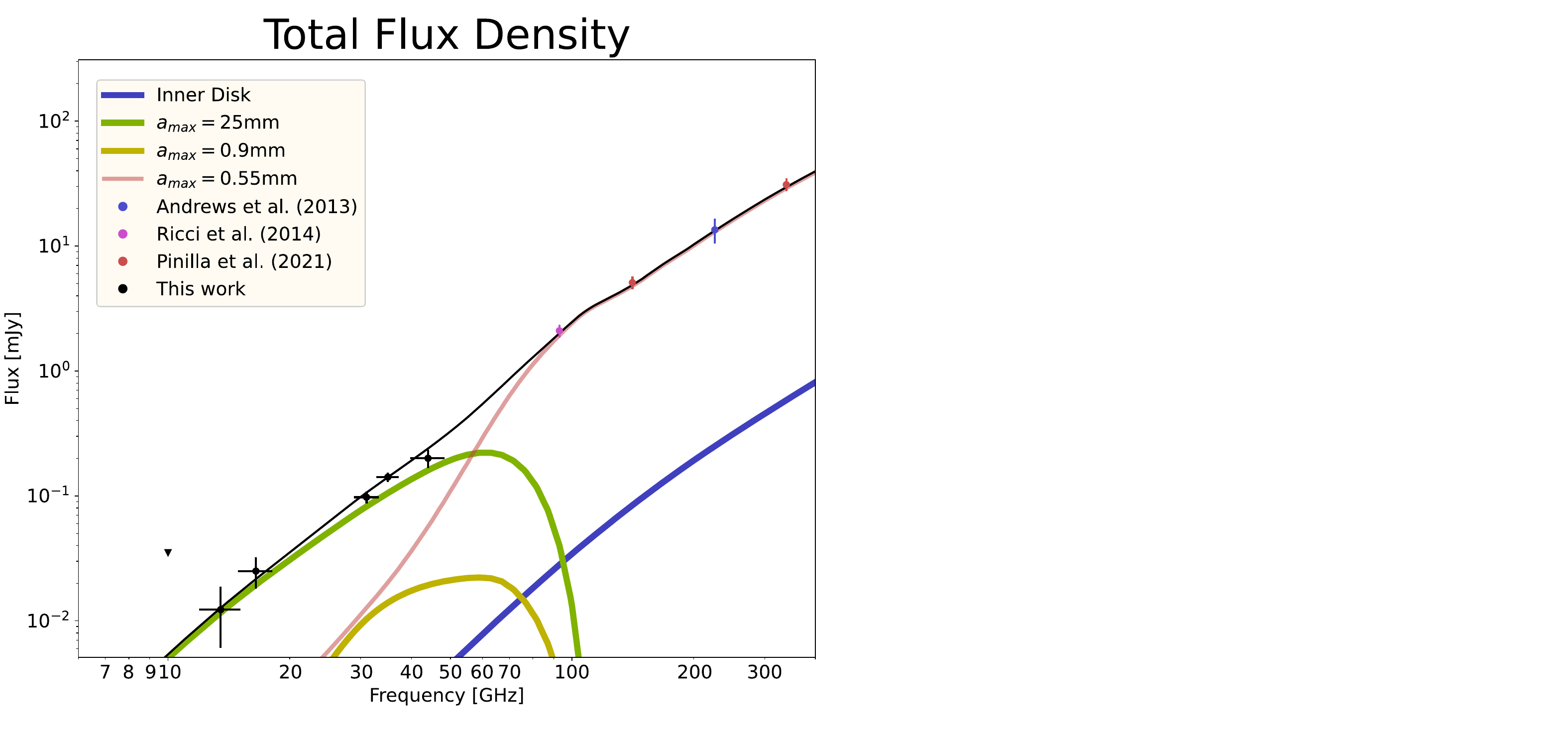} &
         \hspace{-18cm}\includegraphics[height=8.5cm]{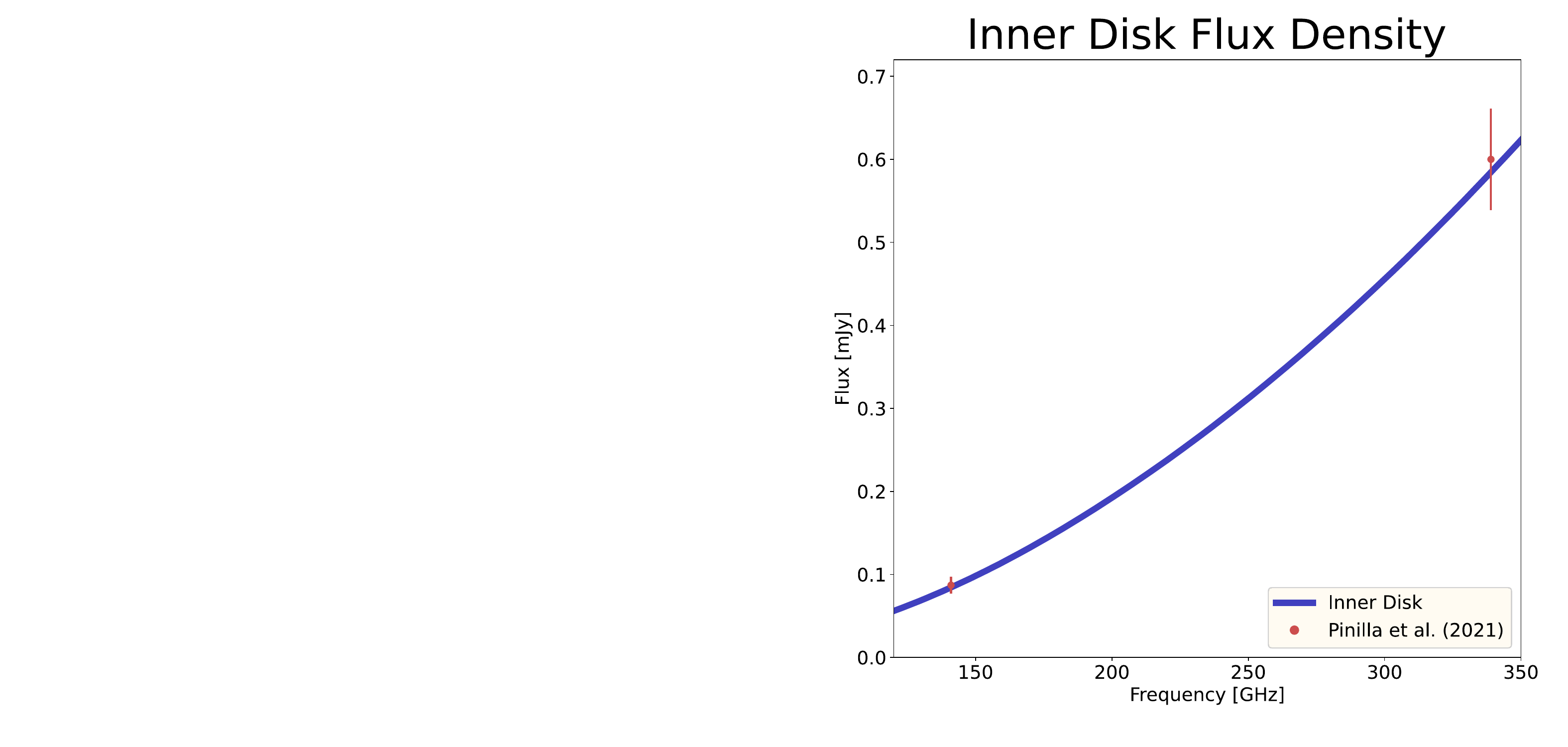} \\
    \end{tabular}
    \caption{
    The observed flux densities from CIDA~1, represented by dots, are compared to our models for interpretation. The flux densities used are listed above the separator line in Table~\ref{tab:fluxdensity}. \jh{The flux uncertainties are explained in the table caption in Table~\ref{tab:fluxdensity}.} Left: The integrated flux densities from CIDA~1 are plotted on a logarithmic scale. Right: The flux density of the CIDA~1 inner disk is shown. In all panels, the blue, green, yellow, and pink lines correspond to our models for the dust emission from the inner disk, an embedded dust ring with $a_{\mbox{\scriptsize max}}=25$~mm, a spatially compact embedded dust component with $a_{\mbox{\scriptsize max}}=0.9$~mm, and an extended dust ring with $a_{\mbox{\scriptsize max}}=0.55$~mm, respectively (refer to Table~\ref{tab:dustmodel} for a summary of our model). The black lines represent the total flux densities of all model components.
    }\label{fig:SED} 
\end{figure*}

\begin{deluxetable}{lccc}[htbp]
\tablecaption{
Flux densities in CIDA 1\tablenotemark{a}\label{tab:fluxdensity}
}
\tablehead{
\colhead{Frequency} &
\colhead{Integrated} &
\colhead{Inner disk} &
\colhead{Refs\tablenotemark{j}}\\
\colhead{(GHz)} &
\colhead{(mJy)} &
\colhead{(mJy)} &
\colhead{}
} 
\startdata
8--12  & $<$0.0351\tablenotemark{b,c} & \nodata & (1) \\
12--15\tablenotemark{d} & \jh{$0.0128\pm0.0061$}\tablenotemark{e} & \nodata & (1) \\
15--18\tablenotemark{d} & \jh{$0.0259\pm0.0067$}\tablenotemark{e} & \nodata & (1) \\
29--33\tablenotemark{d} & \jh{$0.1007\pm0.0095$}\tablenotemark{f} & \nodata & (1) \\
33--37\tablenotemark{d} & \jh{$0.1390\pm0.0097$}\tablenotemark{f} & \nodata & (1) \\
40--48 & \jh{$0.2050\pm0.0350$}\tablenotemark{b,f} & \nodata & (1) \\
93 & $2.1\pm0.21$\tablenotemark{g} & \nodata & (2) \\
141& $5.1\pm0.51$\tablenotemark{g} & $0.087\pm0.0087$\tablenotemark{g,h} & (3) \\
225.5 & $13.5\pm2.8$\tablenotemark{i} & \nodata & (4) \\
339& $31\pm3.1$\tablenotemark{g} & $0.60\pm0.06$\tablenotemark{g,h}& (3) \\ \hline
12--18 & \jh{$0.0190\pm0.0045$}\tablenotemark{b,e} & \nodata & (1) \\
29--37 & \jh{$0.1303\pm0.0074$}\tablenotemark{b,f} & \nodata & (1) 
\enddata
\tablenotetext{a}{Flux densities listed above the separator line are used in our SED modeling in \S~\ref{sub:model}.}\vspace{-0.2cm}
\tablenotetext{b}{Flux densities at \jh{X}, Ku, Ka, and Q-bands without splitting.}\vspace{-0.2cm}
\tablenotetext{c}{The 3~$\sigma$ upper limit.}\vspace{-0.2cm}
\tablenotetext{d}{The Ka and Ku-bands are split into higher and lower bands.}\vspace{-0.2cm}
\tablenotetext{e}{\jh{As the area where the flux is detected at more than 3~$\sigma$ is smaller than the beam size, we regarded the peak flux density as the integrated flux density. The photometric error is the RMS noise in Table~\ref{tab:image}.}}\vspace{-0.2cm}
\tablenotetext{f}{\jh{The integrated flux is measured in the area where the flux is detected at more than 3~$\sigma$. The error is calculated as RMS~$\times \sqrt{N}$, where RMS is taken from Table~\ref{tab:image} and $N$ is the number of beams. As the absolute accuracy of flux in JVLA is 2\% (\url{https://science.nrao.edu/facilities/vla/docs/manuals/oss2017A/performance/fdscale}), we only take account of random noise.}}\vspace{-0.2cm}
\tablenotetext{g}{We assume nominal 10\% (1-$\sigma$) absolute flux uncertainty.}\vspace{-0.2cm}
\tablenotetext{h}{The peak flux density of the central point source.}\vspace{-0.2cm}
\tablenotetext{i}{\jh{The error is taken from \citet{Andrews2013}.}}\vspace{-0.2cm}
\tablenotetext{j}{(1) This study. (2) \citet{Ricci2014a}. (3) \citet{Pinilla2021a}. (4) \citet{Andrews2013}.}\vspace{-0.2cm}
\end{deluxetable}

\section{Constraints on Maximum Grain Sizes} \label{sec:amax}

\subsection{Qualitative interpretation for the SED}\label{sub:sed}

The flux densities measured at all observed frequencies (Table~\ref{tab:fluxdensity}) are primarily attributed to dust emission rather than free-free or synchrotron emissions. If there were strong free-free or synchrotron emission, the intensity at frequencies below 50~GHz would peak at the location of the star itself \citep[e.g., ZZ~Tau~IRS;][]{Hashimoto2022}, which is not detected in Figure~\ref{fig:JVLAimg}. We will discuss later the possibility of spatially extended free-free or synchrotron emission by examining the observed spectral indices.

Weak free-free emission typically exhibits a spectral index close to 0, while synchrotron emission near young stellar objects (YSOs) often shows negative spectral indices \citep[e.g.,][]{Anglada1998Jet}. In contrast, the observed spectral indices at 15--33~GHz and 33--44~GHz are \jh{$2.44\pm0.31$} and \jh{$1.58\pm0.63$}, respectively (calculated using the flux densities from the non-split JVLA bands listed in Table~\ref{tab:fluxdensity}). If there were significant free-free and/or synchrotron emission at frequencies below 50~GHz, we would expect lower spectral indices at lower frequencies. It is also worth noting that free-free and synchrotron emissions are generally very weak in typical Class II YSOs \citep[e.g.,][]{Liu2014ApJ...780..155L,Galvan2014A&A...570L...9G}, so our current findings are not surprising.

The ALMA observations have successfully resolved two sources of dust emission: the inner disk and the dusty ring (refer to Figure~\ref{fig:zoom} in the Appendix). The spectral index of the inner disk at 141--339~GHz is $2.06\pm0.16$ (Table~\ref{tab:fluxdensity}), indicating optically thick dust emission. On the other hand, the dominant flux densities at (sub)millimeter and centimeter wavelengths can be attributed to the dusty ring, which can be further divided into multiple dust emission components through analysis of the spectral energy distribution (SED). Figure~\ref{fig:SED} presents the SED of CIDA~1, which exhibits a change in slope around 40--50~GHz, separating it into the (sub)millimeter region (with a spectral index of \jh{$3.11\pm0.26$} at 44--93~GHz) and the centimeter region (with a spectral index of \jh{$1.58\pm0.63$} at 33--44~GHz). In the (sub)millimeter region, the spectral index remains relatively constant at approximately 2.0 \citep[as shown in Table~\ref{tab:fluxdensity};][]{Ricci2014a,Pinilla2018CIDA1}, whereas it varies with frequency in the centimeter region. These features of the SED require an interpretation that considers dust emission components with different column densities ($\Sigma_{\rm dust}$), temperatures, and maximum grain sizes ($a_{\rm max}$).

The spectral index of \jh{$3.51\pm2.70$} at 13.5--16.5~GHz, derived from splitting the JVLA Ku-band flux densities (Table~\ref{tab:fluxdensity}), provides a qualitative characterization of the upper limit for the maximum grain size ($a_{\rm max}$) of the dominant dust emission source contributing to the detected centimeter flux densities (further discussed in \S~\ref{subsub:ringcm}). However, it is still possible that there are faint dust emission sources with larger $a_{\rm max}$ grains. Through more detailed quantitative modeling, we have determined that the (sub)millimeter region of the spectral energy distribution (SED) is dominated by optically thick dust emission with $a_{\rm max}<1$~mm, while the centimeter region is dominated by dust emission with $a_{\rm max}\gtrsim$1--25 mm (as described in \S~\ref{sub:model}). It should be noted that although \citet{Pinilla2021a} associated the low spectral index of $\sim$2 in the (sub)millimeter region with optically thin emissions from grown dust grains, larger dust grains with $a_{\rm max}>1$~mm are unlikely to explain the steeper spectral index of \jh{$3.11\pm0.26$} at 44--93~GHz presented in Table~\ref{tab:fluxdensity} \citep[refer to, for example, Figure~3 in][]{Draine2006}.

Furthermore, the emission from the dust component with $a_{\rm max}\gtrsim$1--25~mm is suppressed at frequencies above 50~GHz due to obscuration caused by the dust component with $a_{\rm max}<$1~mm. This indicates a vertical segregation of dust grain sizes within the system. A similar vertical segregation of dust sizes was also observed in the previous analysis of the SED for the inner $\sim$10~au region of FU~Ori \citep{Liu2021a}. Our current SED model for CIDA~1, incorporating multiple dust components, shows qualitative similarities to another recently studied very low mass (VLM) star, ZZ~Tau~IRS \citep{Hashimoto2022}. However, in the case of ZZ~Tau~IRS, the SEDs can be explained without assuming mutual obscuration between dust emission components.

\subsection{SED modeling}\label{sub:model}

We conducted SED modeling using the flux densities listed above the separator line in Table~\ref{tab:fluxdensity}. As explained in \S~\ref{sub:sed}, attempting to fit all observed flux densities with a single dust emission slab model would result in a poor fit and misleading parameters. Instead, we followed the approach of \citet{Liu2019c,Liu2021a} and employed a multi-layered dust slab model to fit the integrated (sub)millimeter and centimeter flux densities of CIDA 1 by
\begin{equation}\label{eqn:multicomponent}
    F_{\nu} = \sum\limits_{i} F_{\nu}^{i} e^{-\sum\limits_{j}\tau^{i,j}_{\nu}}, 
\end{equation}
where $F_{\nu}^{i}$ is the flux density of the single dust component $i$ and $\tau^{i,j}_{\nu}$ is the optical depth of the emission component $j$ that obscures the emission component $i$. Since our observations show no evidence of free-free and synchrotron emission (\S~\ref{sub:sed}), we did not include them in our SED models. We assumed that the dusty disk of CIDA 1 is symmetric with respect to the disk midplane, meaning that component~$i$ is closer to the midplane than component~$j$. The dust emission component embedded in the midplane is only obscured by the front side of the other dust emission components. Hence, we assumed that the $\tau^{i,j}_{\nu}$ is half of the total extinction optical depth of dust emission component~$j$. In our models, we considered mutual obscuration only when it was necessary to explain the observations. In cases where the data could be interpreted both with and without assuming mutual obscuration, we assumed there was no obscuration. We will discuss both scenarios qualitatively when relevant.

We determined the flux density and total extinction optical depth of each dust component by utilizing the analytic radiative transfer solutions presented in \citet{Birnstiel2018} (Equations 10--20). These solutions are based on the Eddington approximation. The free parameters for each dust component include the solid angle ($\Omega_{\rm dust}$), dust temperature ($T_{\rm dust}$), dust mass surface density ($\Sigma_{\rm dust}$), and maximum grain size ($a_{\rm max}$). To evaluate the size-averaged, frequency-dependent dust absorption and scattering opacity, we employed the default (i.e., water-ice-coated) DSHARP opacity table from \citet{Birnstiel2018}, which assumes compact dust grains composed of water ice (20~\%), astronomical sillicates (32.91~\%), trollite (7.43~\%), and refractory organics (39.66~\%). It is worth noting that some recent observational studies \citep[e.g.,][]{Ueda2023arXiv230512598U} have raised concerns about the presence of highly porous dust, which is not favored by the DSHARP dust opacity table. For further discussion on this matter, we refer to \citet{Tazaki2019mmPOL}.

In determining the size-averaged dust opacity, we made the assumption that the dust grain size ($a$) distribution, denoted as $n(a)$, follows a power law with an exponent of $-q$ (i.e., $a^{-q}$) between the minimum and maximum grain sizes ($a_{\mbox{\scriptsize min}}$, $a_{\mbox{\scriptsize max}}$), and is zero elsewhere. The size-averaged dust opacity exhibits a very weak dependence on the $a_{\mbox{\scriptsize min}}$, which we set to $10^{-4}$~mm. For our SED models, we adopted a constant dust temperature of $T_{\mbox{\scriptsize dust}}=18.4$~K\footnote{The ring midplane temperature $T_{\rm mid}$ at $r=20$~au is calculated by a simplified expression for a passively heated, flared disk in radiative equilibrium \citep[e.g.,][]{dull2001}: $T_{\rm mid} = \left(\frac{\phi L_{*}}{8\pi r^{2}\sigma_{\rm SB}}\right)^{0.25}$, where $L_{*}$ is the stellar luminosity ($0.19 L_{\odot}$; \citealp{herczeg2014}), $\phi$ is the flaring angle (assumed to 0.02), and $\sigma_{\rm SB}$ is the Stefan--Boltzmann constant.}, except for the unresolved inner disk. The value of 18.4~K represents the median expected dust temperature in the CIDA~1 dust ring. As for the unresolved inner disk, its dust temperature is not constrained by observations and was fixed to 50~K in our SED models to ensure consistency with the Rayleigh-Jeans limit, which is likely the case.

After conducting several trials, we realized that at least four dust components are required to reproduce the frequency-dependent variation of spectral indices: an unresolved inner disk and three components within the dusty ring. These components exhibit qualitative differences in terms of their maximum grain size $a_{\rm max}$ and dust surface density $\Sigma_{\rm dust}$. However, it is important to note that the existing observations do not provide sufficient constraints for the models, as discussed in \S~\ref{subsub:innerdisk}, \S~\ref{subsub:ringcm}, and \S~\ref{subsub:ringmm}). Due to this under-constrained nature of the models, the model parameters are degenerate, making it impossible to perform automatic fits of the model parameters that are meaningful. Instead, we can only iteratively determine plausible model parameters that result in predicted flux densities consistent with the observations. These parameters serve as working hypotheses for the properties of the inner disk and dusty ring. They can be tested and refined through future observations and are summarized in Table~\ref{tab:dustmodel}. While qualitative discussions based on these parameters are feasible, it is important to exercise caution, as some parameters may not have precise face values due to the limited constraints provided by the present observations.

\begin{deluxetable}{ llllll }
\tablecaption{
Dust models for CIDA~1\label{tab:dustmodel}
}
\tablewidth{700pt}
\tabletypesize{\scriptsize}
\tablehead{
\colhead{Component} &
\colhead{$T_{\mbox{\scriptsize dust}}$} &
\colhead{$\Sigma_{\mbox{\scriptsize dust}}$} &
\colhead{$\Omega_{\mbox{\scriptsize dust}}$\tablenotemark{a}} &  
\colhead{$a_{\mbox{\scriptsize max}}$} &
\colhead{Mass} \\
&
\colhead{(K)} &
\colhead{(g\,cm$^{-2}$)} &
\colhead{($10^{-13}$ sr)} & 
\colhead{(mm)} &
\colhead{($10^{-4}$ $M_\sun$)} \\
} 
\startdata
Inner Disk\tablenotemark{b} & 50.0 & 1.5 & 0.055 & 25 & 0.0071\\
$a_{\mbox{\scriptsize max}}$=25 mm\tablenotemark{c} & 18.4 & 20 & 4.2 & 25 & 7.3\\
$a_{\mbox{\scriptsize max}}$=0.9 mm\tablenotemark{c} & 18.4 & 30 & 0.5 & 0.9 & 1.3 \\
$a_{\mbox{\scriptsize max}}$=0.55 mm & 18.4 & 3.0 & 11.5 & 0.55 & 3.0
\enddata
\tablenotetext{a}{1 sr $\sim$4.25$\times$10$^{10}$ square arcsecond, corresponding to $\sim$1.72$\times$10$^{41}$~cm$^2$ at $d=134.6$~pc of CIDA~1  \citep{gaia2016,gaia2022}.}\vspace{-0.2cm}
\tablenotetext{b}{Dust temperature and $a_{\mbox{\scriptsize max}}$ in the inner disk were not constrained by observations (see discussion in \S~\ref{sub:sed}).}\vspace{-0.2cm}
\tablenotetext{c}{Obscured by the $a_{\mbox{\scriptsize max}}=0.55$~mm component which represents the integral dust ring.}
\end{deluxetable}

\subsubsection{Inner disk}\label{subsub:innerdisk}

We approximate the inner disk as an unobscured dust slab with uniform properties. In addition to the dust temperature, three additional free parameters ($\Omega_{\rm dust}$, $\Sigma_{\rm dust}$, $a_{\rm max}$) are required to describe the dust emission within the inner disk. However, due to the limited availability of independent measurements, the model parameters for the inner disk exhibit degeneracy. While it is not possible to determine the physical properties of the inner disk with certainty, modeling its dust emission remains valuable as it enables us to evaluate the contribution of the inner disk emission to the overall integrated flux densities.

We set the maximum grain size $a_{\rm max}$ of the inner disk to the largest value among the components in the dusty ring (25~mm; discussed further in \S~\ref{subsub:ringcm}). The spectral index of the inner disk at 141–339 GHz is $2.20\pm0.16$ (Table~\ref{tab:fluxdensity}), suggesting that the inner disk is marginally optically thick within this frequency range. Consequently, we can only establish a lower limit for its dust surface density $\Sigma_{\rm dust}$, which is approximately 1.5~g~cm$^2$. However, the solid angle ($\Omega_{\rm dust}$) of the inner disk is predominantly determined by the assumed dust temperature, and its face value should not be employed in scientific discussions. It happens that when $a_{\rm max}$ is around 25~mm and the optical depth is very high, the frequency-dependent variation of the dust albedo can account for the observed spectral index of approximately 2.2 at 141--339~GHz (refer to Figure~3 in \citealt{Liu2019SED}). Thus, our current assumption regarding $a_{\rm max}$ serves as a valid and intriguing hypothesis that can be tested through future observations with higher angular resolution at frequencies below 140 GHz. The model indicates that the inner disk makes a negligible contribution to the integrated flux density across all observed frequencies.

\subsubsection{Dusty ring: centimeter region}\label{subsub:ringcm}

The calculated spectral indices for the frequency ranges 13.5--16.5~GHz, 16.5--31~GHz, 31--35~GHz, and 35--44~GHz, based on the flux densities listed above the separator line in Table~\ref{tab:fluxdensity}, are \jh{$3.51\pm2.70$, $2.15\pm0.44$, $2.66\pm0.97$, and $1.70\pm0.81$}, respectively. The drop in spectral indices observed in the frequency ranges of 16.5--31~GHz (12.6~mm) and 35--44~GHz (7.6~mm) can only be explained when considering the presence of multiple dust emission components with distinct values of $a_{\rm max}$ and/or $\Sigma_{\rm dust}$. In fact, to account for the abrupt decrease in spectral indices, it is crucial to consider the effects of dust scattering, as discussed by \citealt{Liu2019SED}. The anomalously lowered spectral indices due to dust scattering occur within a narrow frequency range for a specific $a_{\rm max}$ value. Therefore, although other model parameters may be degenerate, our SED modeling provides a qualitative estimate of $a_{\rm max}$.

In our current fiducial model, we incorporate two dust components with $a_{\rm max}$ values of 25~mm and 0.9~mm (Table~\ref{tab:dustmodel}) to reproduce the flux densities at frequencies below 50~GHz. We assume that there is no mutual obscuration between these two dust components. The spectral index of the $a_{\rm max}=0.9$~mm component is expected to exhibit an anomalously low value around wavelengths of approximately 0.9$\times$2$\pi$~mm ($\sim$53~GHz), which explains the observed low spectral index ($\sim$1.5) at 35--44~GHz. However, the model tends to overestimate the emission at around 44~GHz (Figure~\ref{fig:SED}). It is possible that there are additional dust components with $a_{\rm max}$ values close to 0.9~mm, and they may exhibit complex structures of mutual obscuration. Due to limited signal-to-noise ratio and independent measurements, we are unable to faithfully reproduce these structures. Physically, this suggests the presence of a spatial gradient in $a_{\rm max}$ within a region where $a_{\rm max}$ is around 1~mm. The $a_{\rm max}=0.9$~mm component in our model roughly represents the average dust properties in this region.

Due to the rapid decrease in spectral index of the $a_{\rm max}=0.9$~mm component at frequencies below 30~GHz, the presence of another dust component with a larger $a_{\rm max}$ is required to account for the remaining centimeter emission. However, the determination of the $a_{\rm max}$ for this component is challenging due to the degeneracy between its $\Sigma_{\rm dust}$ and $a_{\rm max}$. Nevertheless, it is evident that the $a_{\rm max}$ of this component must exceed 0.9~mm in order to produce sufficiently strong thermal dust emission at 12--18~GHz. In our current fiducial model, the $a_{\rm max}$ is approximately 25~mm for this component. It is worth noting that it may be possible to resolve the $\Sigma_{\rm dust}$--$a_{\rm max}$ degeneracy and consequently constrain the $a_{\rm max}$ more accurately with follow-up deep JVLA observations at K-band (18--26~GHz). It is also worth considering the possibility that there are additional dust components with larger $a_{\rm max}$, which may exhibit flux densities below the detection threshold at 12--18~GHz in our current observations.

\subsubsection{Dusty ring: (sub)millimeter region}\label{subsub:ringmm}

The spectral index between 44 and 93~GHz is measured to be \jh{$3.11\pm0.26$}, which is significantly higher than the spectral index between 33 and 44~GHz (\jh{$1.58\pm0.63$}). This sudden change in spectral index creates a dip-like feature in the SED around 40--50~GHz (Figure~\ref{fig:SED}). This indicates that the flux density in the 44-93 GHz range is primarily contributed by a dust emission component with an optical depth of approximately unity (marginally optically thick/thin) and a dust opacity spectral index ($\beta$) greater than 1. Consequently, the expected spectral index $\alpha$ is around 3. This dust emission component, with an optical depth of approximately unity, must obscure the $a_{\rm max}=25$~mm and $a_{\rm max}=0.9$~mm components that dominate the centimeter flux densities. Otherwise, the spectral index between 44 and 93~GHz would be too low to align with the observations, considering that the $a_{\rm max}=25$~mm and $a_{\rm max}=0.9$~mm components are bright at frequencies below 50 GHz, possess high column densities (Table~\ref{tab:dustmodel}), and thus have high optical depths (resulting in a spectral index of approximately 2 between 44 and 93~GHz).

The spectral indices between 93--141~GHz, 141--225.5~GHz, and 225.5--339~GHz are measured to be $2.13\pm0.34$, $2.07\pm0.49$, and $2.04\pm0.56$, respectively (Table~\ref{tab:dustmodel}). If we neglect the dust scattering opacity, these spectral indices of approximately 2 over such a wide frequency range can be explained by the modified black body emission, as described by \citet{Hildebrand1983}, of an optically thick dust slab (with an optical depth $\tau\gtrsim10$) in the Rayleigh-Jeans limit. However, considering the assumed dust temperature of 18.4~K, the Rayleigh-Jeans limit ($h\nu\ll k_{b}T$) is not an accurate approximation at $\nu=339$~GHz. The non-Rayleigh-Jeans effect causes the spectral indices to decrease at frequencies above 300~GHz. Moreover, based on our current understanding of dust opacity, it is not physically feasible to simultaneously achieve an optical depth of $\tau\gtrsim10$ at frequencies above 93~GHz and an optical depth of $\tau\sim$1 between 44 and 93~GHz \jh{because dust opacity is not expected to change by more than one order of magnitude in such a narrow frequency range \citep[e.g., Figire~11 of][]{Birnstiel2018}}.

To reproduce the observed spectral indices at frequencies above 93~GHz, we have two options: (1) assuming that a single dust slab with uniform properties (such as $\Sigma_{\rm dust}$ and $a_{\rm max}$) dominates the emission, or (2) considering a mixture of emissions from optically thick and optically thin dust slabs that may have different $a_{\rm max}$ values. Option (1) requires fewer parameters. In this case, we carefully adjust $a_{\rm max}$ and $\Sigma_{\rm dust}$ and rely heavily on the frequency-dependent behavior of dust albedo to match the observed spectral indices. Although it is not always possible to achieve a perfect match using this approach, our study of CIDA~1 demonstrates its feasibility. Specifically, by assuming $a_{\rm max}\sim0.5$~mm, the anomalously lowered spectral indices at $a_{\rm max}\times2\pi$ (as shown in Figure~3 in \citealt{Liu2019SED}) makes it possible to reproduce the spectral indices of approximately 2 when the optical depth is around 1. Additionally, the corresponding anomalously elevated spectral indices at higher frequencies compensate for the non-Rayleigh-Jeans effect.

The alternative approach (case~2) can also explain the observed spectral indices at frequencies above 93~GHz. However, it is generally not preferable to rely on fine-tuning parameters for data interpretation. We find that the dust emission with an $a_{\rm max}$ value of approximately 0.55~mm (as listed in Table~\ref{tab:dustmodel}) is necessary to account for the observed spectral energy distribution (SED) between 44 and 141~GHz. If the $a_{\rm max}\sim0.55$~mm component does not dominate the emission in the (sub)millimeter range, the remaining flux densities in that range may be contributed by dust emission sources that are marginally optically thin and have smaller $a_{\rm max}$ values. Our current model for interpreting the emission above 93~GHz represents an extreme scenario where we mix dust emission sources with various $a_{\rm max}$ values, which in a sense overestimates the average $a_{\rm max}$ of the dust components that dominate the emission above 93~GHz.

\section{Discussions} \label{sec:discussion}

\subsection{Comparison with other sources}\label{sub:Comparison}

Previous analyses of the SED using multi-wavelength data have estimated the maximum grain size $a_{\rm max}$ in the millimeter/centimeter range \citep[e.g.,][]{Carrasco-Gonzalez2019HLTau,Ueda2021HLTau,Macias2021TWHya,Hashimoto2022,Zhang2023HLTau}. \citet{Liu2019SED} highlighted the impact of dust scattering on thermal emission from disks, resulting in anomalously reddened SEDs. This effect has led to an overestimation of $a_{\mbox{\scriptsize max}}$ and an underestimation of dust masses. This finding was further supported by subsequent observational studies conducted by \citet{Ueda2020TWHya}. A comprehensive exploration of parameters using radiative transfer modeling has consistently reached the same conclusion by \citet{Zhu2019massbadget}. \citet{Carrasco-Gonzalez2019HLTau} conducted a re-analysis of the SED of HL~Tau and demonstrated the significant influence of dust scattering in this source as well.

\citet{Ueda2021HLTau} proposed an interpretation for the observed ALMA polarization and SED of HL~Tau. They suggested that millimeter-sized grains settle deeper into the midplane of the disk, while 100~$\mu$m-sized grains remain in the disk atmosphere. This scenario is similar to the case of CIDA~1, where we assume a multi-layered distribution of dust grains (\S~\ref{sec:amax}). In CIDA~1, the emission from larger dust grains with $a_{\rm max}$ values ranging from 1 to 25~mm in the disk midplane is obscured by the presence of small dust grains with an $a_{\rm max}$ of 0.55~mm in the upper layer of the disk.

More recently, \citet{Zhang2023HLTau} explored the porous particles to explain both the polarization and SED of HL~Tau. While polarization observations strongly constrain the $a_{\rm max}$ for compact particles \citep{Kataoka2015a}, porous particles allow for a wider range of $a_{\rm max}$ values that can explain the observed polarization \citep{Tazaki2019mmPOL}. In light of this, \citet{Zhang2023HLTau} proposed the presence of porous grains with $a_{\rm max}$ ranging from 1~mm to 1~m in the HL~Tau disk at $r\lesssim60$~au. To test the potential existence of porous grains in the CIDA~1 ring, conducting multi-wavelength polarization observations with ALMA could prove valuable. Overall, the CIDA~1 ring, with its centimeter-sized grains, represents one of the objects with the largest $a_{\rm max}$ values.

\subsection{Dust mass}

The dust mass of the CIDA 1 ring is approximately 383~$M_\earth$ according to Table~\ref{tab:dustmodel}. Comparing it with previous ALMA studies \citep[e.g.,][]{Andrews2020ARAA}, the dust mass of the CIDA~1 ring appears to be exceptionally large. One possible reason for our higher estimate of the CIDA~1 ring's mass is that we assumed it to be optically thick. \citet{Zhu2019massbadget} discussed that a compact disk ($<30$~au), where most of the disk is optically thick in ALMA (sub-)millimeter observations, can underestimate the actual dust mass by a factor of 10. While \citet{Pinilla2021a} estimated a dust mass (including the central point source) of approximately 10~$M_\earth$ for the CIDA~1 ring, assuming it to be optically thin, our estimate using the optically thick model for grains with $a_{\rm max}=$~0.55~mm, which ALMA is sensitive to, is around 100~$M_\earth$. These differences could explain the disparity. Similar to our estimation, SED analyses based on multi-wavelength data of TW~Hya suggested a dust mass approximately five times higher than those derived from (sub-)millimeter disk surveys \citep{Macias2021TWHya}.

\subsection{Gravitational stability}\label{sub:GI}
Assuming a gas-to-dust mass ratio of the canonical value of 100 \citep{bohlin78}, the total disk mass (gas~$+$~dust) of CIDA~1 is approximately 0.12~$M_\sun$ (Table~\ref{tab:dustmodel}), while CIDA~1 itself has a mass of 0.19~$M_\sun$ \citep{Kurtovic2021a}. A large disk-to-star mass ratio exceeding 0.1 can potentially lead to gravitational instability \citep[e.g.,][]{kratter16}. We will now examine the gravitational stability of the system. Theoretical studies \citep[e.g.,][]{durisen2007} suggest that if the \citet{toomre1964} Q-parameter, defined as  
\begin{eqnarray}
Q=\frac{c_{\rm s} \Omega_{\rm K}}{\pi G \Sigma_{\rm disk}},
\end{eqnarray}
where $c_{\rm s}$, $\Omega_{\rm K}$, and $\Sigma_{\rm disk}$ represent the sound speed, the Keplerian angular velocity, and the disk's gas surface density, respectively, is less than unity, the disk may be prone to gravitational instability. With a midplane temperature of 18.4~K at the ring located at $r=20$~au (refer to \S~\ref{sub:sed}) and a gas surface density of $5.3\times10^3$~g/cm$^{-2}$ (Table~\ref{tab:dustmodel}), assuming the gas-to-dust mass ratio of 100, the Toomre's Q-parameter at the CIDA~1 ring is approximately 0.06.

A Q-parameter well below unity indicates that the disk is strongly unstable, in which spiral arms are expected to be present \citep[e.g.,][]{dong15gi}. However, CIDA~1 does not exhibit such spiral arms in its disk \citep{Pinilla2021a,Kurtovic2021a}. Estimating the Q-value comes with a major uncertainty, namely the gas-to-dust mass ratio. If the ring represents a dust trap, the local gas-to-dust ratio could be significantly lower than 100. To further constrain on the gas mass in the CIDA~1 system, observations of gas-mass tracers such as HD \citep{Bergin2013HD} would provide valuable insights.

\subsection{Speculation of planet formation around CIDA~1}\label{sub:implication}

CIDA~1, with the mass of 0.19~$M_\sun$ \citep{Kurtovic2021a}, falls into the category of very low-mass (VLM) stars, typically defined as having a mass of $\lesssim0.2 M_\sun$. According to the core (pebble) accretion scenario, these systems are expected to be capable of forming only low-mass planets, typically ranging up to a few Earth masses \citep{Payne2007a,Ormel2017a,Schoonenberg2019,liu2019a,Liu2020a,miguel2020a,burn2021}. 

According to the pebble accretion scenario \citep[e.g.,][]{liu2019a}, the growth of planetary cores is impeded when a growing core creates a partial gas gap in the disk, because the local gas pressure bump efficiently traps accreting pebbles \citep[e.g.,][]{Pinilla2012dusttrap}. Consequently, the planet becomes isolated from further pebble accretion, and its mass is referred to as the `pebble isolation mass' ($M_{\rm iso}$). The $M_{\rm iso}$ is determined by the central stellar mass and the disk aspect ratio ($h_{\rm g}=h/r$) and is given by the following relationship: $M_{\rm iso}\approx25(h_{\rm g}/0.05)^3(M_{\star}/M_\sun)M_\earth$ \citep[e.g.,][]{liu2019a}. In flared disks, where the $h_{\rm g}$ is generally higher in outer disk regions \citep[e.g.,][]{Hayashi1985PPII}, the $M_{\rm iso}$ can reach masses comparable to giant planets, even for VLM stars.

For CIDA 1, the $h_{\rm g}$ value at the ring position of $r=20$~au is estimated to be 0.076\footnote{The gas scale height $h$ is the ratio of the gas sound speed to the angular velocity ($h=c_s/\Omega$), then the disk aspect ratio can be written as $h_{\rm g} = h/r \approx 0.03 (\frac{M_{\rm star}}{M_\sun})^{-\frac{1}{2}} (\frac{T_{\rm mid}}{300{\rm K}})^{\frac{1}{2}} (\frac{r}{1{\rm au}})^{\frac{1}{2}}$, where $T_{\rm mid}$ is a disk midplane temperature with 18.4~K (\S~\ref{sub:model}).}, resulting in an $M_{\rm iso}$ of 16.7~$M_\earth$. The dust mass in the CIDA~1 ring is estimated to be approximately 383~$M_\earth$ (Table~\ref{tab:dustmodel}), which could provide sufficient material for the formation of giant planets \citep{Jang2022}. The presence of centimeter-sized pebbles in the CIDA~1 ring (\S~\ref{sec:amax}) may further facilitate the growth of planetary cores \citep[e.g.,][]{Morbidelli2015}. Therefore, it is plausible that giant planets could form within the CIDA~1 ring during the typical disk lifetime of a few million years.

\jh{Another avenue of planet formation in the CIDA~1 ring is gravitational instability \citep[e.g.,][]{Boss1997GI,Mercer2020}. When the gravitationally-unstable disks cool sufficiently fast ($t_{\rm cool}\Omega\lesssim$~a~few, where $t_{\rm cool}$ and $\Omega$ are the cooling time and orbital frequency), they may fragment \citep[e.g.,][]{Gammie2001GI,Rice2003GI,Dong2016GI}, resulting in bound, self-gravitating objects (i.e., companions). As discussed in \S~\ref{sub:GI}, the CIDA~1 system might be gravitationally unstable but the disk does not show eminent asymmetric structures. More studies, especially on gas-mass estimates, are needed for further investigation. 
}   
 
\section{Conclusion} \label{sec:conclusion}

We conducted observations of the CIDA~1 ring using the JVLA at centimeter wavelengths ranging from $\lambda=0.7$~cm \jh{(44~GHz)} to 3.0~cm \jh{10~GHz}. We successfully detected signals at $\lambda=0.7$~cm \jh{(44~GHz)}, 0.9~cm \jh{(33~GHz)}, and 2.0~cm \jh{(15~GHz)}. At $\lambda=0.7$ \jh{(44~GHz)} and 0.9~cm \jh{(33~GHz)}, we were able to spatially resolve a partial ring structure. Although a potential weak asymmetry was observed in the ring, further deep observations are required to achieve a more reliable detection. Based on the spatial distributions and spectral indexes of these emissions, we attributed them to dust rather than free-free or synchrotron emissions.

The emission in the frequency range of 15--340~GHz was analyzed using a simple SED model consisting of four dust components. It is important to note that our modeling results depend on the uncertain properties of the dust. The model includes an inner dust disk component, as well as dust ring components with maximum grain sizes ($a_{\mbox{\scriptsize max}}$) of 25, 0.9, and 0.55~mm. The analysis suggests that grain growth in the CIDA~1 ring proceeds to at least centimeter-sized grains.

While recent ALMA dust polarization observations have suggested the presence of sub-millimeter-sized dust grains (approximately 100~$\mu$m) in Class~II disks \citep[e.g.,][]{Kataoka2016a}, it is important to note that these observations were biased towards bright disks. Despite being faint in comparison (with a total flux of approximately 10~mJy at $\nu=225.5$~GHz; \citealp{Andrews2013}) compared to the brighter DSHARP disks ($\gtrsim100$~mJy at band 6; \citealp{andr2018}), the CIDA~1 ring, with its centimeter-sized grains, is placed in a unique parameter space in terms of dust size. The exact origins of grain growth in the CIDA~1 ring remain uncertain. Expanding the sample size to include more disks with millimeter to centimeter-sized grains will provide valuable insights into grain growth and shed light on key physical processes such as the snowline, the orbital timescale, and turbulence.

\begin{acknowledgments}

The authors thank the anonymous referee for a timely and constructive report. 
We thank Paola Pinilla for sharing fits files of ALMA dust continuum images at band-4 and 7. 
The National Radio Astronomy Observatory is a facility of the National Science Foundation, operated under a cooperative agreement by Associated Universities, Inc.
This work has made use of data from the European Space Agency (ESA) mission {\it Gaia} (\url{https://www.cosmos.esa.int/gaia}), processed by the {\it Gaia} Data Processing and Analysis Consortium (DPAC, \url{https://www.cosmos.esa.int/web/gaia/dpac/consortium}). Funding for the DPAC has been provided by national institutions, in particular the institutions participating in the {\it Gaia} Multilateral Agreement.
This study used the following ALMA data: ADS/JAO. ALMA \#2011.0.00259.S and 2018.1.00536.S. 
ALMA is a partnership of ESO (representing its member states), NSF (USA), and NINS (Japan), along with NRC (Canada), MOST and ASIAA (Taiwan), and KASI (Republic of Korea), in cooperation with the Republic of Chile. The Joint ALMA Observatory is operated by ESO, AUI/NRAO, and NAOJ.
This study was supported by JSPS KAKENHI Grant Numbers 21H00059, 22H01274, 23K03463 and 18H05441.  T.M. is supported by Yamada Science Foundation Overseas Research Support Program.

\end{acknowledgments}

\software{
          astropy \citep{2013A&A...558A..33A},  
          Numpy \citep{VanDerWalt2011}, 
          CASA \citep[v6.4.3; ][]{mcmu07},
          }

\facilities{JVLA, ALMA}

\appendix

\section{Zoomed-in Ka-band image}\label{secA:zoom}

The image with the highest signal-to-noise ratio (SNR) among our observations, captured in the Ka-band, is overlaid on the contours of the ALMA band-7 image for a closer view.

\begin{figure}[htbp]
         \includegraphics[width=\linewidth]{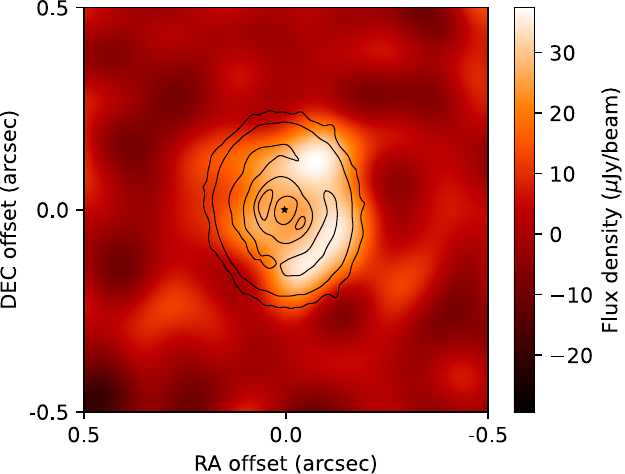} 
    \caption{
    Zoomed-in Ka-band image (0$\farcs$18~$\times$~0$\farcs$16) superposed with contours of ALMA band-7 image \citep[54~mas~$\times$~37~mas;][]{Pinilla2021a}. Contour levels are 17.1~$\mu$Jy\,beam$^{-1}$ (1~$\sigma$) $\times$[5, 15, 45]. The stellar position is marked by a black star
    }\label{fig:zoom}
\end{figure}

\bibliography{sample63}{}
\bibliographystyle{aasjournal}

\end{document}